\input harvmac
\def\np#1#2#3{Nucl. Phys. B {#1} (#2) #3}
\def\pl#1#2#3{Phys. Lett. B {#1} (#2) #3}
\def\plb#1#2#3{Phys. Lett. B {#1} (#2) #3}
\def\prl#1#2#3{Phys. Rev. Lett. {#1} (#2) #3}
\def\physrev#1#2#3{Phys. Rev. D {#1} (#2) #3}

\def\semidirect{\mathbin{\hbox{\hskip2pt\vrule height 4.1pt depth -.3pt
width .25pt \hskip-2pt$\times$}}}
\def\ev#1{\langle#1\rangle}

\def\O{{\cal O}}
\def\Ot{{\cal O}_{\tau}^{(-4)}}
\def\Otbar{{\cal O}_{\overline \tau}^{(4)}} 
\def\f#1#2{\textstyle{#1\over #2}}
\def\lfm#1{\medskip\noindent\item{#1}}
\lref\malda{J.M. Maldacena, hep-th/9711200, Adv. Theor. Math. Phys. 2
(1998) 231.}
\lref\GKP{S.S. Gubser, I. Klebanov, and A.M. Polyakov, hep-th/9802109,
\pl{428}{1998}{105}.}
\lref\EW{E. Witten, hep-th/9802150, Adv. Theor. Math. Phys. 2 
(1998) 253.}
\lref\kleba{I. Klebanov, hep-th/9702076, \np{496}{1987}{231}.}
\lref\KR{R. Kallosh and A. Rajaraman, hep-th/9805041,
\physrev{58}{1998}{125003}.}
\lref\JS{J. Schwarz and P. West, \pl{126}{1983}{301}; P. Howe and 
P. West, \np{238}{1984}{181}; J. Schwarz, \np{226}{1983}{269}.}
\lref\GSW{M. Green, J. Schwarz, and E. Witten, {\it Superstring
Theory}, Cambridge University Press, 1987.}
\lref\GMZ{M. Gunaydin, D. Minic, M. Zagermann, hep-th/9806042,
\np{534}{1998}{96}.}
\lref\KS{S. Kachru and E. Silverstein, hep-th/9802183,
\prl{80}{1998}{4855}.}
\lref\LNV{A. Lawrence, N. Nekrasov, and C. Vafa, hep-th/9803015,
\np{533}{1998}{199}.}
\lref\KW{I. Klebanov and E. Witten, hep-th/9807080,
\np{536}{1998}{199}.} 
\lref\sugrp{P.G.O. Freund and I. Kaplansky, Jour. Math. Phys. 17
(1976) 228; M. Scheunert, W. Nahm, and V. Rittenberg,
Jour. Math. Phys. 17 (1976) p. 1626; V. Kac, Adv. in Math. 26 (1977)
8.}
\def\adsr{\refs{\malda, \GKP, \EW}}
\def\sugra{\refs{\JS}}
\def\sdet{{\rm sdet}}
\def\d{\delta}
\lref\witten{E. Witten, hep-th/9803131, Adv. Theor. Math. Phys. 2
(1998) 505.}
\lref\ferr{L. Andrianopoli and S. Ferrara, hep-th/9803171,
\pl{430}{1998}{248};  S. Ferrara, M.A. Lledo, and A. Zaffaroni,
hep-th/9805082, \physrev{58}{1998}{105029}.}
\lref\GM{M. Gunaydin and N. Marcus, Class. Quant. Grav. 2, (1985) L11.}
\lref\LMRS{S. Lee, S. Minwalla, M. Rangamani, and N. Seiberg,
hep-th/9806074.}
\lref\BG{T. Banks and M. Green, hep-th/9804170. JHEP 9805 (1998) 002.}
\lref\DFS{E. D'Hoker, D.Z. Freedman, and W. Skiba, hep-th/9807098, 
\physrev{59}{1999}{045008}.}
\lref\GG{M.B. Green and M. Gutperle, hep-th/9701093,
\np{498}{1997}{195}.} 
\lref\SPA{S.P. de Alwis, hep-th/9607011; \pl{388}{1996}{291}.}
\lref\FMMR{D.Z. Freedman, S.D. Mathur, A. Matusis, and L. Rastelli,
hep-th/9804058, \np{546}{1999}{96}.}
\lref\JP{J. Polchinski, hep-th/9510017, \prl{75}{1995}{4724}.}
\lref\HMM{J. Harvey, R. Minasian, and G. Moore, hep-th/9808060.}
\lref\HS{M. Henningson and K. Skenderis, hep-th/9806087.}
\lref\SW{L. Susskind and E. Witten, hep-th/9805114.}
\lref\GKPeet{S.S. Gubser, I.R. Klebanov, A.W. Peet, 
hep-th/9602135, \physrev{54}{1996}{3915}.}
\lref\BHE{I.R. Klebanov and A.A. Tseytlin, hep-th/9604089,
\np{475}{1996}{164}; S.S. Gubser and I.R. Klebanov, hep-th/9708005,
\pl{413}{1997}{164}.}
\lref\PPN{M. Pernici, K. Pilch, and P. van Nieuwenhuizen,
\pl{143}{1984}{103}.} 
\lref\PNT{K. Pilch, P. van Nieuwenhuizen, and P.K. Townsend,
\np{242}{1984}{377}.}
\lref\KRN{H.L Kim, L.J. Romans, and P. van Nieuwenhuizen,
\physrev{32}{1985}{389}.}
\lref\harmrefs{P Howe and P. West, hep-th/9509140; hep-th/9607060,
\pl{389}{1996}{273}; hep-th/9611074; hep-th/9808162,
\pl{449}{1998}{341}.}
\lref\HWinvts{P.S. Howe and P.C. West, hep-th/9611075,
\pl{400}{1997}{307}.} 
\lref\AFGJ{D. Anselmi, D.Z. Freedman, M.T. Grisaru, A.A. Johansen,
hep-th/9708042, \np{B526}{1998}{543}.}
\lref\actref{E. Bergshoeff, B. Janssen, and T. Ortin,
hep-th/9506156, Class. Quant. Grav. 13 (1996) 321.}
\lref\sixteen{N. Seiberg, hep-th/9705117, Nucl. Phys. Proc. Suppl. 67
(1998) 158.}
\lref\HST{P.S. Howe, K.S. Stelle, P.K. Townsend, \np{191}{1981}{445}.}
\lref\aneesh{A.V. Manohar, hep-th/9802419, 1997 Les Houches proc. to
appear.}
\lref\FPZ{S. Ferrara, M. Porrati, A. Zaffaroni, hep-th/9810063.}
\lref\cornwell{J.F. Cornwell, {\it Group Theory in Physics, vol. 3},
Academic Press, 1989.}
\lref\GS{M.B. Green and S. Sethi, hep-th/9808061.}
\lref\BSS{L. Brink, J. H. Schwarz, and J. Scherk, \np{121}{1977}{77}.}
\lref\NOS{M. Rocek and W. Siegel, \pl{105}{1981}{275}.}
\lref\BGKR{M. Bianchi, M. Green, S. Kovacs, G. Rossi, hep-th/9807033,
JHEP 9808 (1998) 13.}
\lref\KPoo{A. Kehagias, H. Partouche, hep-th/9710023,
\pl{422}{1998}{109}; hep-th/9712164 Int. J. Mod. Phys. A13 (1998) 
5075.}
\lref\RMAT{R.R. Metsaev, A.A. Tseytlin, hep-th/9805028,
\np{533}{1998}{109}; hep-th/9806095, \pl{436}{1998}{281}.}
\lref\PST{P. Pasti, D. Sorokin, M. Tonin, hep-th/9611100,
\physrev{55}{1997}{6292}.}
\lref\AFGJ{D. Anselmi, D.Z. Freedman, M.T.Grisaru, and A.A.
Johansen, hep-th/9608125, \plb{394}{1997}{329}; hep-th/9708042,
\np{526}{1998}{543}.}  
\lref\GRPS{F. Gonzalez-Rey, I. Park, K. Schalm, hep-th/9811155, 
\pl{448}{1999}{37}.}
\lref\EHSSW{E. Eden, P.S. Howe, C. Schubert, E. Sokatchev, P.C. West, 
hep-th/9811172.}
\lref\DZF{D.Z. Freedman, private communication.}
\lref\DHKMV{N. Dorey, T.J. Hollowood, V.V. Khoze, M.P. Mattis, and
S. Vandoren, hep-th/9901128.}

\Title{hep-th/9811047,  UCSD/PTH-98/37, IASSNS-HEP-98/92}
{\vbox{\centerline{Bonus Symmetries of ${\cal N}=4$ Super-Yang-Mills} 
\centerline{Correlation Functions via AdS Duality
}}}
\medskip
\centerline{Kenneth Intriligator}
\vglue .5cm
\centerline{UCSD Physics Department}
\centerline{9500 Gilman Drive}
\centerline{La Jolla, CA 92093}
\vglue .25cm
\centerline{and}
\vglue .25cm
\centerline{School of Natural Sciences\footnote{${}^*$}{address for
Fall term, 1998.}}
\centerline{Institute for Advanced Study}
\centerline{Princeton, NJ 08540, USA}

\bigskip
\noindent

General conjectures about the $SL(2,Z)$ modular transformation
properties of ${\cal N}=4$ super-Yang-Mills correlation functions are
presented.  It is shown how these modular transformation properties
arise {}from the conjectured duality with $IIB$ string theory on
$AdS_5\times S^5$.  We discuss in detail a prediction of the AdS
duality: that ${\cal N}=4$ field theory, in an appropriate limit, must
exhibit bonus symmetries, corresponding to the enhanced symmetries of
$IIB$ string theory in its supergravity limit.

\Date{11/98}           

\newsec{Introduction and summary}

As with all conjectured dualities, that of \malda\ between ${\cal N}=4$
supersymmetric $SU(N)$ Yang-Mills and $IIB$ string theory with $N$
units of $F_5$ flux, which compactifies on $AdS_5\times S^5$, relates
the weakly coupled limit of one theory to the strongly coupled limit
of the dual.  The string side is weakly coupled in the limit of small
$g_s=4\pi g_{YM}^2$ and large 't Hooft coupling $\lambda
\equiv g_{YM}^2N$ \refs{\kleba, \malda}, where it can be approximated
by semi-classical $IIB$ supergravity. In this limit, the field theory
dual is strongly coupled, as the relevant coupling is $\lambda
=g_{YM}^2N$, and perturbation theory is not valid.  The mapping
between weak coupling of one theory and strong coupling of the dual
makes duality very powerful, but also difficult to check unless one
has independent, non-perturbative information about at least one of the
dual theories.  

A first non-trivial check of the duality
\malda\ is that both theories have the same symmetry group, 
$PSU(2,2|4)$, which has bosonic subgroup $SU(2,2)\times SU(4)_R$ and
32 supercharges.  Also, both have the $SL(2,Z)$ $S$-duality group
\malda.  $PSU(2,2|4)$ has short representations (to be discussed in
detail in what follows), labeled by positive integers $p$, whose
$SU(2,2)\times SU(4)_R$ quantum numbers are completely fixed in terms
of $p$ and thus not renormalized.  In the ${\cal N}=4$ gauge theory,
the independent $p$ are the degrees of the Casimirs of the gauge
group.  In the dual $IIB$ supergravity on $AdS_5\times S^5$, $p$
corresponds to the $S^5$ Kaluza-Klein spherical harmonics of massless
10d supergravity fields
\refs{\GKP, \EW}.  The two sides, the spectrum of short representation
operators in the $4d$ field theory, versus KK modes in the 5d $AdS$
supergravity, agree in the large $N$ limit \EW.

Non-renormalization theorems are known for a few ${\cal N}=4$ field
theory current correlation functions, which can thus be used to check
the conjectured duality.  More generally, the feeling is that the
power of ${\cal N}=4$ supersymmetry has not been fully exploited and
that there are other non-renormalization theorems waiting to be
discovered.  Quantities for which the answer {}from weakly coupled
gravity differs {}from that of weakly coupled field theory presumably
do not satisfy a non-renormalization theorem (assuming the duality is
correct) and the answer {}from weakly coupled gravity is regarded as a
non-trivial {\it prediction} for strongly coupled field theory.

It sometimes happens that the weakly coupled gravity result
unexpectedly agrees with that of free field theory; this can be
regarded as evidence for a new non-renormalization theorem.  This was
the case in the results of \LMRS\ for three-point functions of
normalized primary operators in short multiplets.  This led the
authors of \LMRS\ to conjecture that these 3-point functions are
independent of the 't Hooft coupling in the large $N$ limit and
perhaps even independent of $g_{YM}$ for arbitrary $N$.  The fate of
the CFT/AdS correspondence is completely independent of the fate of
such a conjectured non-renormalization theorem; nevertheless, the
latter is an interesting question in the field theory.  Evidence for
the conjectured non-renormalization of such three-point functions of
primary operators was obtained in \DFS, where it was shown in a purely
field theory analysis for small $g_{YM}$ that, for all $N$, leading
order radiative corrections to all such two-point and three-point
correlation functions surprisingly conspire to cancel.  This possibly
hints at a larger symmetry of the ${\cal N}=4$ theory.

We discuss predictions for such a larger symmetry of ${\cal N}=4$
field theory based on assuming the duality with $IIB$ string theory on
$AdS_5\times S^5$.  In the limit where $IIB$ string theory is
approximated by $IIB$ supergravity, there are additional approximate
symmetries: the $SL(2,Z)$ symmetry is enlarged to an $SL(2,R)$
symmetry and there is its maximal compact subgroup, $U(1)_Y$, which
enters into the description of interacting $IIB$ supergravity in terms
of an $SL(2,R)/U(1)_Y$ coset.  These enhanced approximate symmetries
must then also show up in the dual ${\cal N}=4$ gauge theory in the 
appropriate limit.

Stringy corrections to $IIB$ supergravity, which generally violate
these approximate enhanced symmetries, are suppressed when
\eqn\strexp{{\alpha '\over L^2}\sim {1\over \sqrt{g_{YM}^2N}}\ll 1.}
Here $L$ is the size of both $AdS _5$ and $S^5$, which is related by
flux quantization to the units $N$ of $F_5$ flux by
\eqn\Lis{{L^4\over \kappa _{10}}\sim N,} 
with $\kappa _{10}$ the 10d gravitational coupling.  

The condition 
\strexp\ alone is not sufficient to ensure that stringy corrections
are suppressed, as $D$-string effects
also lead to $SL(2,R)$ and $U(1)_Y$ violating terms; to have these
effects also be suppressed, we also need
\eqn\strexpp{{\widetilde \alpha '\over L^2}\sim \sqrt{{g_{YM}^2\over
N}}\ll 1.}  It is in the double limit, where both \strexp\ and
\strexpp\ are satisfied, that our bonus symmetries of ${\cal N}=4$ Yang
Mills theories are predicted to hold; in what follows, we will
refer to this as the ``double limit.''  Clearly the double
limit requires large $N$.   Because the natural, dimensionless, 
quantum expansion parameter of the gravity dual is 
\eqn\hbaris{\hbar \sim {\kappa _5^2\over
L^3}\sim {\kappa _{10}^2\over L^8}\sim N^{-2},}
where $\kappa _5$ is the 5d gravitational coupling, which is related
to $\kappa _{10}$ by dimensional reduction, 
$\hbar \ll 1$ and the gravity dual is
semi-classical in the double limit.  
\def\double{\strexp\ and \strexpp}

It must be stressed that the larger symmetry applies {\it
only} to those operators of ${\cal N}=4$ Yang-Mills which correspond
to states in supergravity.  Those operators in long multiplets which
correspond to stringy states, which are expected to have large
anomalous dimension $\Delta \sim (g_{YM}^2N)^{1/4}$ in the double
limit \adsr, should not be expected to respect these symmetries.  We
consider here only operators in the standard short multiplets of
$PSU(2,2|4)$; these always correspond to states visible in
supergravity.  The bonus symmetry of the double limit should also
extend to those operators in long multiplets which map to non-stringy,
multi-particle supergravity states\foot{I am grateful to N. Seiberg
for reminding me about these long multiplets.}, though this will not
be discussed here.

We consider, then, arbitrary correlation functions of operators
$\O_i(x)$ in short representations of the superconformal group:
\eqn\gencor{\ev{\prod _{i=1}^n\O_i(x_i)}=f_{i_1\dots i_n}(x_i;N; 
g_{YM}, \theta _{YM}).}  We argue that a prediction of the duality of
\adsr\ is that, in the double limit discussed above, 
the leading behavior of all such correlation functions
is \eqn\gencorp{\ev{\prod _{i=1}^n\O _i(x_i)}=N^2f_{i_1\dots
i_n}(x_i),} where the functions are independent of $N$ and $g_{YM}$
and $\theta$ to leading order.  The $N$ dependence, as will be
discussed, is associated with tree-level supergravity.  The reason for
the $g_{YM}$ and $\theta$ independence of \gencorp\ is the $SL(2,R)$
symmetry of supergravity: because $SL(2,R)$ maps the gauge coupling
\eqn\taumap{\tau\equiv {\theta _{YM}\over 2\pi}+{4\pi i\over
g_{YM}^2}\rightarrow {a\tau +b\over c\tau +d}, \qquad \pmatrix{a&b\cr
c&d}\in SL(2,R),} which can be used to map any $\tau$ in the
upper-half-plane to any other $\tau$, correlation functions in this
limit must be independent of $\tau$.  For arbitrary correlation
functions of operators in short multiplets, the leading term in the
double limit is thus predicted to be always completely independent of
the 't Hooft coupling $\lambda =g_{YM}^2N$! Because $SL(2,R)$ is
broken to $SL(2,Z)$ in the full string theory, correlation functions
are generally expected to have non-trivial $\tau$ dependence in the
terms which are sub-leading in the double limit.  The normalization of
the operators $\O _i$, which is important in making sense of the
statement \gencorp, will be discussed in the next section.

It is also interesting to consider the local $U(1)_Y$, which is the
maximal compact subgroup of $SL(2,R)$, and enters in the
$SL(2,R)/U(1)_Y$ description of $IIB$ supergravity, which is briefly
reviewed in sect. 3.  Although $U(1)_Y$ is a local symmetry, there is
no corresponding gauge field and thus no corresponding conserved
current in the field theory.  Nevertheless, $U(1)_Y$ leads to a
non-trivial $R$-type symmetry, under which the super-charges
transform, of the superconformal algebra.  It is non-trivial that the
superconformal algebra admits such a symmetry, as will be discussed in
sect. 4.  The operators $\O _i(x)$ in short representations of the
superconformal group can all be assigned definite charges, opposite to
those of the supergravity fields to which these operators couple.  The
$U(1)_Y$ symmetry of supergravity implies a selection rule for field
theory correlation functions of operators
\eqn\Ycons{\ev{\prod _{i=1}^n\O _i^{(q_i)}(x_i)}=0
\qquad\rm{unless}\quad \sum _{i=1}^nq_i=0,}
where $\O _i^{(q_i)}$ is a short-multiplet operator of $U(1)_Y$ charge
$q_i$.  As we will discuss in sect. 5, $U(1)_Y$ is {\it not} a
symmetry of the field theory; nevertheless, it it is predicted to
yield approximate selection rules \Ycons\ in the double limit
of \strexp\ and \strexpp.

The $\tau$ independence of \gencorp\ actually follows as a consequence
of the selection rule \Ycons.  To see this, note that the derivative
of an arbitrary $n$-point correlation function with respect to the
gauge coupling $\tau$ is given by
\eqn\corrd{\partial _\tau \ev{\prod _i\O _i(x_i)}=\tau _2^{-1}
\int d^4z\ev{
\Ot (z)\prod _i\O _i(x_i)},}
where $\Ot $ is the exactly marginal operator, to be discussed in
detail in what follows, which couples to $\tau$ in the action; it's
the on-shell ${\cal N}=4$ Lagrangian.  There is a conjugate operator
$\Otbar$ which couples to $\overline \tau$, allowing us to
independently vary both $g_{YM}$ and $\theta$.  The $U(1)_Y$ charge of
$\Ot$ is $-4$, as indicated by the superscript.  It follows {}from
\Ycons\ and \corrd\ that non-zero correlation functions are
independent of $\tau$, as in \gencorp.  

In sect. 6 we make some general conjectures about the $SL(2,Z)$
modular transformation properties of ${\cal N}=4$ super-Yang-Mills
correlation functions.  For any gauge group, we conjecture that
arbitrary correlation functions transform under $SL(2,Z)$ modular
transformations as
\eqn\genslz{\ev{\prod _i \O ^{(q_i)}_i(x_i)}_{{a \tau +b\over c\tau
+d}}= \left({c\overline \tau +d\over c\tau +d}\right) ^{q_T/4}
\ev{\prod _i \O ^{(q_i)}_i(x_i)}_{\tau},}
with $q_T=\sum _i q_i$ the net $U(1)_Y$ charge of the correlation
function.  (In the case of $Sp(n)$ and $SO(2n+1)$, which are exchanged
by $\tau \rightarrow -1/\tau$, the correlation functions on the two
sides of \genslz\ would be for these two dual groups; because we are
only discussing $SU(N)$, this will not concern us here.)
In the supergravity limit, where $SL(2,Z)$ is extended to
$SL(2,R)$, the transformation \genslz\ implies the $\tau$ independence
of \gencorp\ and the $U(1)_Y$ selection rule \Ycons.  

String theory leads to higher dimension terms in the effective action
which violate the $SL(2,R)$ and $U(1)_Y$ symmetries.  This agrees with
the fact that these are not symmetries of ${\cal N}=4$ field theory
for general $g_{YM}$ and $N$.  The predicted form of the corresponding
corrections to the ${\cal N}=4$ field theory correlation functions,
away {}from the double limit, is discussed in sect. 7.  These
corrections, which violate $SL(2,R)$ and $U(1)_Y$, are subleading by
at least $N^{-3/2}$, for fixed $g_{YM}$, and satisfy our $SL(2,Z)$
modular transformation rule \genslz.  For small $g_{YM}$, these
corrections are sub-leading by at least order $(g_{YM}^2N)^{-3/2}$.

Based on the form of the stringy violations of $U(1)_Y$ found in the
$\alpha '$ expansion of $IIB$ string theory, we conjecture that
$U(1)_Y$ is actually an exact symmetry of $n\leq 4$-point functions,
i.e. valid for all $g_{YM}$ and $N$. Using \corrd, this would have as
a consequence the exact $SL(2,R)$ invariance of $n\leq 3$-point
functions, in line with the conjecture and calculations of
\refs{\LMRS,
\DFS}.  In sect. 8 we discuss some aspects of attempting to prove
exact $U(1)_Y$ invariance of $n$-point functions with low $n$, though
we only succeeded in finding a simple proof of exact $U(1)_Y$
invariance for $n=2$-point functions.  The exact $U(1)_Y$ invariance
of 2-point functions implies that arbitrary $n$-point functions also
respect $U(1)_Y$ in the leading Born-approximation appropriate for
{\it small} $g_{YM}^2N$.

In sect. 9 we examine $U(1)_Y$ in the context of the ${\cal N}=4$
harmonic superspace formalism of \harmrefs, and find a contradiction:
assuming the validity of this formalism and the classification of
invariants in \HWinvts, we prove that an {\it arbitrary} $n$-point
correlation function would {\it exactly} respect $U(1)_Y$, for all
$g_{YM}$ and $N$, for any $n$.  This result would imply that all
$n$-point correlation functions of operators in short multiplets would
be exactly independent of $g_{YM}$ for all $g_{YM}$ and $N$, a result
which is definitely\foot{In the original version of this paper, the
conclusion that arbitrary $n$-point functions are not renormalized was
referred to as ``highly suspicious,'' and it was pointed out that it
could probably be disproved directly in perturbation theory by
generalizing the calculations of \DFS\ to 4-point functions.  It was
also pointed out that such non-renormalization would already be in
conflict with the analysis of \BGKR, where it was shown that
Yang-Mills instantons do contribute to certain four and higher-point
correlation functions.  Subsequently it was pointed out to me by
D. Freedman \DZF\ that the four-point function of the stress tensor
$T_{\mu \nu}$ {\it must} get renormalized, already in perturbation
theory, because of results already appearing in
\AFGJ : the OPE of two $T_{\mu \nu}$ stress tensors contains the
Konishi current, and the anomalous dimension of the Konishi current
receives $g_{YM}$ quantum corrections (even in the ${\cal N}=4$
theory).  In addition, the first-order radiative contributions to the
four-point function of the superconformal primary operator $\O _2$ (to
be discussed in what follows) were subsequently explicitly calculated
\refs{\GRPS, \EHSSW}\ and were indeed found to be non-vanishing.  In
sum, the result we obtained via ${\cal N}=4$ harmonic superspace is
definitely incorrect.} {\it incorrect} for general $n$-point
functions!  As discussed further in sect. 9, this contradiction shows
that the ${\cal N}=4$ harmonic superspace formalism is either invalid
or incomplete.  This issue does not in any way affect the results or
conclusions of the other sections of this paper.
 
The enhanced approximate $SL(2,R)$ and $U(1)_Y$ symmetries of 
the double limit \double\  are also predicted to occur in the
${\cal N}=2,1,0$ Yang-Mills theories associated with orbifolds of the
${\cal N}=4$ theory \refs{\KS,\LNV} and with the ${\cal N}=1$ theory
of \KW.  They should also occur for the $3d$ ${\cal N}=0$ theory
obtained {}from the 4d theory at finite temperature.  However, there is
no analog of these additional global symmetry in the case of 11d
supergravity (which has no symmetries), so no such enhanced
approximate symmetry is to be expected for the $3d$ or $6d$ theories
associated via \malda\ with $M$ theory on $AdS_4\times S^7$ or
$AdS_7\times S^4$.  Therefore, the $4d$ ${\cal N}=0$ theory obtained
as in \witten, {}from a compactification of the 6d theory which breaks
supersymmetry will also not have such enhanced approximate symmetries.

The $U(1)_Y$ symmetry also entered in the discussion in a recent work
on ${\cal N}=6$ supergravity and $SU(2,2|3)$ superconformal invariance
\FPZ, which appeared in the final stages of writing up this paper.  In
particular, the discussion in the last section of \FPZ\ has some
overlap with the bonus symmetries discussed here.

\newsec{The normalization of ${\cal N}=4$ operators}

Before discussing the enhanced symmetries of supergravity, we here
consider some basic points concerning the $N$ dependence of
correlation functions of operators in ${\cal N}=4$ $SU(N)$ gauge
theory in the large $N$ limit.  The primary operators $\O _p$ of small
representations of the superconformal group are Lorentz scalars, with
dimension $\Delta =p$, and in the $SU(4)_R$ representation with Dynkin
indices $(0,p,0)$ (corresponding to a Young tableaux with $p$ columns,
each two rows deep).  In terms of the $SU(N)$ adjoint scalar $\phi$,
which is in the $(0,1,0)$ (i.e. ${\bf 6}$) representation of the
$SU(4)_R$ global symmetry, $\O _p\sim [\Tr _{SU(N)} (\phi
^p)]_{(0,p,0)}$; the subscript means to keep only the $(0,p,0)$
representation, which is obtained by taking the totally symmetric,
traceless product of the $p$ $\phi$'s.

There is a normalization of the operators $\O _p$ which is natural for
the large $N$ limit and convenient for comparing with supergravity.
We start with the fields normalized so that the ${\cal N}=4$ gauge
theory lagrangian is
\eqn\lagN{{\cal L}=N\cdot {1\over g_{YM}^2N }(-\f{1}{4}\Tr F_{\mu
\nu}^2+\half (D\phi)^2+\overline \psi D\!\!\!\!\slash\psi +\dots).}
We then normalize the $\O _p$ as
\eqn\opnorm{\O _p=N(g_{YM}^2N)^{-p/2}[\Tr _{SU(N)}(\phi
^p)]_{(0,p,0)}.}  

A virtue of this normalization can be seen in terms
of the rescaled fields $\widehat \phi=\phi /\sqrt{g_{YM}^2N}$, with
sources introduced for the composite operators:
\eqn\lagnsource{{\cal L}=N\cdot (\half \Tr (D\widehat \phi )^2+\dots +
\sum _pJ_p\Tr (\widehat \phi ^p)).}
The overall factor of $N$ in \lagnsource\ simplifies the $N$-counting:
for arbitrary sources $J_p$, the connected vacuum graph with Euler
character $\chi =2-2g-b$ is of order 
\eqn\corrsource{S_{eff}^{\rm{field theory}}[J_p]\sim N^\chi}
in the large $N$ limit; see e.g. \aneesh.  The leading contribution in
the large $N$ limit comes {}from planar diagrams and is of order
$N^2$.  In terms of the original fields $\phi$ entering
\lagN, the normalization of the operator coupling to the source $J_p$ 
in \lagnsource\ is that of  \opnorm.  Thus arbitrary correlation
functions of the operators normalized as in \opnorm\ satisfy
\eqn\borncor{\ev{\prod _{i=1}^n\O _{p_i}(x_i)}=N^2f_{p_i}(x_i; \lambda
\equiv g_{YM}^2N)} 
in the planar limit.

The factor of $g_{YM}^{-p}$ in \opnorm\ ensures that in \borncor\ the
functions $f_{p_i}(x_i;\lambda )\rightarrow f_{p_i}(x_i)$ are 
independent of $\lambda$ in the free-field, Born approximation
appropriate for $\lambda \rightarrow 0$.  As will be discussed in
sect. 6, these factors of $g_{YM}^{-p}$ are also crucial for ensuring
nice $SL(2,Z)$ modular transformation properties of the operators and
correlation functions.

Having fixed the normalization of the primary operators $\O _p$ as in
\opnorm, the normalization of all other operators in the short
superconformal multiplet, which are descendents of $\O_p$, are fixed
by acting with the $Q$ and $\overline Q$ (the structure of the
multiplet will be discussed in detail in what follows).  Thus all
operators in the small representation of the superconformal group
labeled by $p$ have the same $N(g_{YM}^2N)^{-p/2}$ normalization as in
\opnorm.  The most general correlation function of all such operators
then has the same $N^2$ dependence as in \borncor\ in the large $N$
limit, and the same independence of $\lambda$ in the $\lambda
\rightarrow 0$ limit.

According to the prescription in \refs{\GKP, \EW}\ for computing
${\cal N}=4$ correlation functions via the duality of \malda,
\corrsource\ is understood as the supergravity or string theory 
effective action with the boundary condition that the fields equal the
sources $J_p(x)$ on the boundary of $AdS_5$.  The above normalization
of the operators and sources nicely agrees with this method of
computation.  This is because the quantum loop expansion parameter,
$\hbar$, of the supergravity or string theory dual is given by
\hbaris.  The $g$ loop contribution to the effective action with
fields set to equal the sources $J_p(x)$ on the boundary of $AdS_5$ is
thus given by
\eqn\ssugran{S_{eff}^{\rm{gravity}}[J_p]\sim \hbar ^{1-g}\sim N^{\chi},}
with $\chi=2-2g$, exactly as in \corrsource; in particular, the
leading, semi-classical contribution to $S_{eff}$ is $\sim N^2$.
Normalizing the operators as in \opnorm\ corresponds to normalizing
the supergravity fields, which approach the sources $J_p$ on the
boundary, without any unnatural factors of $\hbar$.

\newsec{Review of 
the $U(1)_Y$ and $SL(2,R)$ symmetries of $IIB$ supergravity}

It is perhaps useful to briefly review some textbook (see, e.g. \GSW)
facts about $IIB$ supergravity.  Type $IIB$ supergravity in 10d has a
$U(1)$ symmetry which rotates the two chiral supersymmetries, and thus
is an $R$ symmetry, which we will refer to as $U(1)_Y$.  Normalizing
the supercharges to have $U(1)_Y$ charge $\pm 1$, the complex scalar
dilaton has $Y=4$, the complex two-form gauge field $B_{\mu
\nu}$ has $Y=2$, the complex
Weyl spinor dilatino $\lambda$ has $Y=3$, and the complex Weyl
gravitino $\psi$ has $Y=1$.  The complex conjugate fields have the
opposite $Y$ charges and the remaining fields, which are real, all
have $Y=0$
\sugra.  The entire collection of massless physical fields can be
described in terms of a 10 superfield $\Phi (x,\theta)$, where the
Grassmann coordinate $\theta$ is in the complex Weyl ${\bf 16}$ of
$SO(9,1)$ and $\Phi$ is subject to the constraint $\overline D\Phi =0$
and also $D^4\Phi =\overline D^4\overline \Phi =0$.  

The interacting $IIB$ supergravity theory is formulated in terms of a
$SL(2,R)/U(1)_Y$ coset.  Originally, for convenience, the coset was
given in terms of $SU(1,1)\cong SL(2,R)$ \sugra; the $SL(2,R)$ form
can be found e.g. in \GG\ and will be briefly reviewed here.  The
scalars are given in terms of the ``driebein'' field, which is used 
to convert between $SL(2,R)$ indices $\alpha = 1,2$ and 
$U(1)$ charges $Y=\pm 2$ 
\eqn\Vmatrix{V=(V^\alpha _-,V^\alpha _+)=(-2i\tau _2 
)^{-1/2}\pmatrix{\overline \tau e^{-i\phi}&\tau e^{i\phi}\cr
e^{-i\phi} &e^{i\phi}};} this $V$ is related to an element of
$SL(2,R)$ by a change of basis to a complex basis.  $V$ transforms
under global $SL(2,R)$ and local $U(1)_Y$ transformations as $V^\alpha
_\pm
\rightarrow e^{\pm 2i\Sigma (x)} U^\alpha _\beta V^\beta _\pm$, where
$U^\alpha _\beta \in SL(2,R)$, $\Sigma$ is the local $U(1)$ phase and
the normalization reflects our choice of normalizing $V^\alpha _\pm$
to have $U(1)_Y$ charge $\pm 2$.  The real scalar $\phi$ in \Vmatrix\
is unphysical and can be set to zero by choice of $U(1)_Y$ gauge.
$SL(2,R)$ transformations only preserve the gauge fixed form of $V$
when accompanied by particular $U(1)_Y$ transformations and the upshot
is that $\tau$ in \Vmatrix\ has the standard transformation \taumap\
under $SL(2,R)$.  

The driebein $V^\alpha _\pm$ is used to convert all other fields to
be invariant under $SL(2,R)$ but charged under $U(1)_Y$; so the
dilaton $\tau$ will be the only field to transform under
$SL(2,R)$.  In particular, the complex antisymmetric tensor $A_{\mu
\nu}^\alpha$ (with $A_{\mu \nu}^1=A_{\mu \nu}^2{}^*$), which
transforms as an $SL(2,R)$ doublet and neutral under $U(1)_Y$, is
converted to the $SL(2,R)$ singlet $B_{\mu \nu}=\epsilon _{\alpha
\beta} V_+^\alpha A_{\mu \nu}^\beta$, which has $U(1)_Y$ charge 2,
as in the free theory spectrum mentioned above.  The
$SL(2,R)$ invariant object 
\eqn\Pis{P_\mu=-\epsilon _{\alpha\beta}V^\alpha
_+\partial _\mu V_+^\beta ={i\over 2\tau _2}\partial _\mu \tau} 
has $U(1)_Y$ charge 4, in line with the $U(1)_Y$ charge of the dilaton
of the free theory mentioned above.  Similarly, the remaining fields
and $U(1)_Y$ charges are as mentioned above for the free theory, and
are all $SL(2,R)$ singlets.  

Given the principle that supersymmetry should respect the
$SU(1,1)\cong SL(2,R)$ and $U(1)_Y$ symmetries, with the supercharges
carrying charge $\pm 1$ under $U(1)_Y$, it was shown in \sugra\ that
consistency of the super-algebra completely determines (actually over
determines) the form of the supersymmetry variations up to a single,
real, dimensionful coupling constant $\kappa$, which is the 10d
gravitational coupling constant.  Finally, requiring closure of this
super-algebra determines the interacting $IIB$ supergravity equations
of motion
\sugra, as the algebra only closes on shell. The equations of motion
determined in this way will clearly also respect the $SL(2,R)$ and
$U(1)_Y$ symmetries.  Even in the gauge fixed form, with the
unphysical degree of freedom in $V$ eliminated, the equations of
motion found in \JS\ manifestly respect a residual global $U(1)_Y$
symmetry, under which the fields have the charge assignments given
above.  

In converting the discussion of \JS\ to one in which $SL(2,R)$ is used
instead of $SU(1,1)$, there is a small subtlety with regard to the
$U(1)_Y$ symmetry.  In the $SU(1,1)$ formulation, the $SU(1,1)$
invariant object \Pis\ is given upon gauge fixing $U(1)_Y$ 
by $P_\mu =(1-B^*B)^{-1}\partial _\mu B$ and, since $P_\mu$ is
assigned $U(1)_Y$ charge $4$, the complex scalar $B$ also 
carries $U(1)_Y$ charge 4.  In the $SL(2,R)$ form \Pis, $P_\mu$ again
has $U(1)_Y$ charge 4, but $\tau$ does not have a well-defined
$U(1)_Y$ charge assignment because of the $\tau _2$ in \Pis.  The
reason is that the map between $B$ and $\tau$  
\eqn\btmap{\tau =i{1-B\over 1+B}}
maps the origin $B=0$, where $U(1)_Y$ is unbroken, to $\tau =i$ and a
simple $U(1)$ phase for $B$ gives a more complicated transformation
for $\tau$.  More generally, non-zero $\ev{B}$ or $\ev{\tau}$
spontaneously break $U(1)_Y$.  For our purposes, however, it is useful
to note that the leading order variation $\delta \tau $ of $\tau$
around a constant $\ev{\tau }$ can be assigned a well-defined $U(1)_Y$
charge. As \Pis\ gives $P_\mu =i \partial _\mu \delta\tau /2\ev{\tau
_2} $, we can assign $U(1)_Y$ charge 4 to $\delta \tau$ and zero to
$\ev{\tau _2}$.  In any case, $SL(2,R)$ invariance implies that
amplitudes expanded around vanishing fields and constant $\ev{\tau}$
will be independent of $\ev{\tau}$.  For this reason, the spontaneous
breaking of $U(1)_Y$ by $\ev{\tau}$ will not be relevant for our
concerns.

The action which gives the equations of motion, modulo the
self-duality of $F_5$ which can be imposed by hand or treated as in 
\PST, takes the
$SL(2,R)$ and $U(1)_Y$ invariant form in the Einstein frame \actref:
\eqn\slinvact{S={1\over 2\kappa _{10}^2}\int
d^{10}x\left[\sqrt{-g}(R- {\partial \tau \partial \overline
\tau\over 2\tau _2^2}-\f{1}{12}G^{\mu \nu\lambda}G^*_{\mu \nu \lambda}
-\f{1}{4\cdot 5!}F_5^2)
-\f{1}{(12)^3}C_4\wedge G\wedge
G^*+\cdots\right] ,}
with $G=\epsilon ^{\alpha \beta}V_{+\alpha}dA_{2,\beta}=\tau
_2^{-1/2}(\tau d B_2+dC_2)$ and $F_5=dC_4+5\epsilon ^{\alpha
\beta}A_{2,\alpha }\wedge dA_{2, \beta}$.  

The $SL(2,R)$ and $U(1)_Y$ symmetries of $IIB$ supergravity will be
respected by all tree-level amplitudes, and thus by the generating
functional of these tree level amplitudes.  

\newsec{Representations of the superconformal group 
$PSU(2,2|4)$ and its $U(1)_Y$ automorphism}

Because $F_5$ is neutral under $SL(2,R)$ and $U(1)_Y$, they will also
be symmetries of the supergravity theory with $N$ units of $F_5$ flux
and vacuum $AdS_5\times S_5$.  In particular, $U(1)_Y$ must act as an
$R$-symmetry of the superconformal group $PSU(2,2|4)$.  It is
non-trivial that $PSU(2,2|4)$ indeed does admit such an outer
automorphism.

In order to clarify the connection between the $U(1)_Y$ of
supergravity and the superconformal group, it is useful to review a
general subtlety of the supergroups $SU(M|N)$ when $M=N$; see e.g.
\cornwell\ for useful facts about super 
matrices, groups, and algebras.  Our case of interest is $M=(2,2)$ and
$N=4$; the non-compact signature of $M$ will not introduce any further
subtleties.  An element of the $u(M|N)$ algebra can be written as
\eqn\sumnalg{g=\pmatrix{A&B\cr C&D},}
with $A\in u(M)$ and $D\in u(N)$ bosonic and $B$ and $C$ fermionic.
There is a decoupled $u(1)_D$ ideal generated by $g_D=1_{M+N}$ and an
$R$-symmetry $u(1)_Y$ generated by
\eqn\gy{g_Y=\half \pmatrix{1_M&0\cr 0& -1_N},}
under which $A$ and $D$ are neutral and the generators $B$ and $C$
have charge $\pm 1$.  For $M\neq N$, the ideal $u(1)_D$ is eliminated
by the condition $str g\equiv \tr A-\tr D=0$; the resulting algebra is
$su(M|N)$, which contains $u(1)_R$ generated by $g_R=g_Y+\half
(M+N)(N-M)^{-1}g_D$ in its bosonic subalgebra.  On the other hand, for
$M=N$ the condition $str g=0$ eliminates $u(1)_Y$
\gy\ rather than $g_D=1_{N+N}$ 
and thus $su(N|N)=psu(N|N)\oplus u(1)_D$ does not contain the
$R$-symmetry generated by $u(1)_Y$.
Although $u(1)_Y$ is not contained in $su(N|N)$ or $psu(N|N)$, it
clearly acts as a consistent automorphism on them: indeed, these
groups can be consistently extended to include $g_Y$
\gy\ as an additional element by simply not imposing the $str g=0$
condition.  The larger group thus obtained, which we refer to as
$PU(N|N)$ in the case where the decoupled $U(1)_D$ is eliminated by
hand, is $U(1)_Y\semidirect PSU(N|N)$, rather than $U(1)_Y\otimes
PSU(N|N)$, since $U(1)_Y$ acts as a non-trivial $R$ symmetry on the
fermionic generators.

Representations of $PSU(2,2|4)$ can be assigned definite charges under
the $U(1)_Y$ automorphism group.  The short representations of
$PSU(2,2|4)$ were constructed by the oscillator method in \GM.  The
full short representation is labeled by an integer $p>0$ and consists
of a number of particular representations of the bosonic
$SU(2,2)\times SU(4)_R$ subgroup.  The motivation in \GM\ was to use
$PSU(2,2|4)$ representation theory to understand the spectrum of
fields in 5d, ${\cal N}=8$ supergravity; the same spectrum was
obtained as with linearized KK reduction of $IIB$ supergravity on
$S^5\times AdS_5$ in \KRN, where $p$ is related to the KK spherical
harmonic.  The relation between these supergravity fields and
operators in the 4d ${\cal N}=4$ gauge theory was discussed in detail
in \EW\ and the fact that these operators are also classified by the
$PSU(2,2|4)$ representation theory of \GM\ was emphasized in
\ferr.  

In addition to finding the $SU(2,2)\times SU(4)_R$ quantum numbers, the
$U(1)_Y$ charges of the representations were also determined in \GM,
where it was appreciated that the 5d $N=8$ supergravity must also have
the $U(1)_Y$ symmetry of the 10d $IIB$ supergravity.  The $U(1)_Y$
charges of the 5d supergravity fields are simply those of the
corresponding 10d $IIB$ supergravity field of which the 5d field is a
$S^5$ spherical harmonic KK mode.  We emphasize again that $U(1)_Y$
acts as a non-trivial $R$-symmetry on $PSU(2,2|4)$; clearly $U(1)_Y$
of supergravity is an $R$-symmetry since the graviton is neutral and
the gravitino is charged.  This differs {}from a brief discussion in
\GMZ, where the $U(1)_Y$ of supergravity was instead identified with
the decoupled, non-R-symmetry $u(1)_D$ in $su(N|N)=u(1)_D\oplus
psu(N|N)$.

For convenience, we included the table of representations and $U(1)_Y$
charges determined in \GM\ in appendix $A$.  We changed the
normalization of the $U(1)_Y$ charges for convenience and also changed
the signs to be those of the operators in the ${\cal N}=4$ field
theory, which are of exactly opposite $U(1)_Y$ charge {}from the
supergravity fields to which these operators couple. Also indicated in
the table is the supermultiplet form of the representations: there is
a primary representation $\O _p$, which has $U(1)_Y$ charge 0, and
superconformal descendents obtained by acting with powers of the
supercharges $Q_\alpha ^I$ and $\overline Q_{I, \dot
\alpha}$ on $\O _p$, with $Q$ represented by $\d$ and $\overline Q$
represented by $\overline \d$.  When the representations $\O _p$ are
operators rather than fields, it should be understood that the
$\delta^r\overline \d ^s\O _p$ appearing in the table is shorthand for
a nested sequence of commutators and anti-commutators with the
supercharges, e.g. $\d ^2\overline \d\O _p$ should be understood as
$[Q, \{Q,[\overline Q,\O _p]\}]$.  The supercharge descendent
structure truncates at $\d ^4\overline \d ^4\O _p$ rather than at $\d
^8\overline \d ^8\O _p$ because it is a short rather than long
$PSU(2,2|4)$ representation.  A representation $\d ^r\overline \d ^s\O
_p$ has $U(1)_Y$ charge $s-r$.

The representations with $p<4$ truncate further.  The $SU(4)_R$ quantum
numbers of the representations are given by the Dynkin labels
$(l_1,l_2,l_3)$ (which corresponds to a Young tableaux with $l_k$
columns of boxes which are $k$ rows deep, $k=1,2,3$).  Those
representations which would have Dynkin index $l_2<0$ according to the
table, of course, vanish.  The $p=0$ representation contains the
identity as its only element and the $p=1$ representation is the
decoupled representation sometimes referred to as the singleton or
doubleton; it is not present if the ${\cal N}=4$ Yang-Mills group is
simple.  The $p=2$ representation is the ``massless'' representation
which contains, among other operators, the conserved superconformal
currents.  

Another interesting pair of operators in the $p=2$ representation are
\eqn\otauis{\Ot=\d ^4\O_2\qquad\rm{and}\qquad\Otbar=
\overline \d ^4\O _2.}
These operators are Lorentz and $SU(4)_R$ singlets, and exactly
marginal as $\Delta =4$.  They are also annihilated when acted on with
any more powers of $Q$ or $\overline Q$ since all such descendents in
the short representation would have a $SU(4)_R$ Dynkin index $l_2<0$
for $p=2$.  In the ${\cal N}=4$ gauge theory, $\Ot$ is the exactly
marginal operator corresponding to changing the gauge coupling $\tau$.
The corresponding field in supergravity to which $\Ot$ couples is the
lowest KK mode of the dilaton, which we also denote by $\tau$, which
has $U(1)_Y$ charge $+4$. $\Ot$ in the gauge theory will be discussed
further in the next section.

\newsec{${\cal N}=4$ gauge theory and the $U(1)_Y$ non-symmetry}

There are some points to be made concerning how $U(1)_Y$ acts in the
${\cal N}=4$ gauge theory.  To illustrate a first point, it will
suffice to consider Abelian $U(1)$ ${\cal N}=4$ gauge theory.  The
fields are the gauge field $A_{\alpha
\dot \alpha}$, 
scalars satisfying the reality condition $\phi ^{[IJ]}\equiv
(\phi_{[IJ]})^*=\half \epsilon ^{IJKL}\phi _{[KL]}$, and fermions
$\psi _{I,\alpha}$, $\overline \psi ^I_{\dot \alpha}$, where the $I$
is a fundamental $SU(4)_R$ representation index.  The on-shell
supersymmetry transformations are given by
\eqn\susyvar{\eqalign{\delta A_{\alpha \dot \alpha}&= \overline \eta
^{I\dot \beta}\epsilon _{\dot \alpha \dot \beta} \psi _{I\alpha}+\eta
^\beta _I\epsilon _{\alpha \beta}\overline \psi ^I_{\dot \alpha}\cr
\delta \phi _{[IJ]}&=\eta ^\alpha _{[I}\psi _{J]\alpha}+\epsilon
_{IJKL} \overline \eta ^{K\dot \alpha}\overline \psi ^L_{\dot
\alpha}\cr 
\delta \psi _{I\alpha}&=\eta ^\beta _IF_{(\alpha \beta )}+\overline
\eta ^{J\dot \beta}\partial _{\alpha \dot \beta}\phi _{IJ}\cr
\delta F_{(\alpha \beta )}&=\overline \eta ^{I\dot \gamma }\partial
_{\dot \gamma (\alpha }\psi _{\beta )I},}} where $\eta ^\alpha _I$ and
$\overline \eta ^{I\dot \alpha}$ are Grassmann parameters to keep
track of the action of $Q_{\alpha}^I$ and $\overline Q _{I \dot
\alpha}$, there are similar transformations for 
$\overline \psi ^l_{\dot
\alpha}$ and $\overline F_{\dot \alpha \dot \beta}$, 
and we have left out numerical constants for simplicity.  (This
notation differs {}from that of the appendix, where $\delta$ denotes
acting with $Q_\alpha ^I$ only.) There is no known off-shell
formulation\foot{Note that the on-shell amplitudes in the supergravity
or string theory dual apparently do provide a fully supersymmetric,
off-shell formulation of the ${\cal N}=4$ superconformal symmetry of
the boundary field theory.}  of ${\cal N}=4$ supersymmetry at $\phi
=0$ \NOS.

The fields transform under $U(1)_Y$ with the charges
\eqn\charges{\matrix{&&
\phi _{IJ}&\psi _{I\alpha} &\overline \psi ^I _{\dot\alpha}
& F_{(\alpha \beta )}&\overline F_{(\dot \alpha \dot \beta )}\cr
U(1)_Y&& 0&-1&1&-2&2}.}  Note that this transformation is compatible
with the $\phi$ reality condition, because $\phi$ is neutral, but
bizarre, because $F_{\alpha \beta}$ is not neutral. It is not the same
as the $U(1)$ in $U(4)\cong U(1)\times SU(4)_R$.  Assigning charges $1$
to $\eta _I^\alpha$ and $-1$ to $\overline \eta ^{I\dot \alpha}$, this
transformation is respected by all of the supersymmetry variations
\susyvar\ with the exception of that of $A_{\alpha \dot \alpha}$,
which is not a gauge invariant physical field anyway.  The $U(1)_Y$
transformation is also a symmetry of the equations of motion for the
physical fields.  Indeed, as the super-transformations \susyvar\
are purely on-shell, they close on the equations of motion, which must
then also respect $U(1)_Y$.  Note that $F_{\alpha \beta}$ involves
$\vec E+i\vec B$, and thus the $U(1)_Y$ symmetry involves a continuous
rotation between electric and magnetic fields - i.e. a continuous
version of the discrete electric-magnetic duality transformation $S$.

While $U(1)_Y$ is a symmetry of the equations of motion of the Abelian
theory, it is not a symmetry of the lagrangian:
\eqn\abact{\eqalign{{\cal L}&=\tau (-\f{1}{4}F_{\alpha \beta}F^{\alpha
\beta} +\half \overline
\psi ^I_{\dot \alpha}\partial ^{\alpha \dot \alpha}\psi
_{I\alpha}-\f{i}{2}\partial _{\alpha \dot\alpha}\phi ^{IJ}\partial
^{\alpha \dot \alpha}\phi _{IJ})\cr
&+\overline \tau (-\f{1}{4}\overline F_{\dot \alpha \dot
\beta}\overline F_{\dot
\alpha \dot \beta}
-\half \overline
\psi ^I_{\dot \alpha}\partial ^{\alpha \dot \alpha}\psi
_{I\alpha}+\f{i}{2}\partial _{\alpha \dot\alpha}\phi ^{IJ}\partial
^{\alpha \dot \alpha}\phi _{IJ}).\cr}} It is trivially a symmetry if
the equations of motion are imposed, as then the lagrangian simply
vanishes.  The subtlety of having to impose the equations of motion is
also apparent in our identification of $\Ot =\d ^4\O _2$ as the exactly
marginal, supersymmetry preserving, operator corresponding to changing
$\tau$.  Applying $Q^4$ using
\susyvar\ to $\O _2\sim \phi _{[IJ]}\phi _{[KL]}-(trace)$ 
gives $\d ^4\O _2\sim F^2$, which is zero upon imposing the equations
of motion.  This corresponds to varying $\tau$ in the lagrangian with
the equations of motion imposed; this is trivial in the Abelian case.

In the non-Abelian case, the supersymmetry transformations \BSS\ are 
\susyvar\ modified by 
replacing all $\partial _{\alpha\dot \alpha}\rightarrow D_{\alpha \dot
\alpha}$ gauge covariant derivatives and there is an additional term
in
\eqn\susyvara{\delta \psi _{I\alpha}=\dots +\eta ^\beta _J\epsilon
_{\alpha \beta}[\phi _{IK},\phi ^{JK}].}  Assigning charges \charges\
and charge $+1$ to $\eta ^\beta _J$ as before, we see that the
additional term \susyvara\ does not respect the $U(1)_Y$ symmetry.
Thus $U(1)_Y$ is {\it not} a symmetry for general $g_{YM}$ and $N$.
When combined with the operation of changing the sign of all fields, a
$Z_4$ subgroup of $U(1)_Y$ is preserved, but uninteresting, as it is
simply the center of the $SU(4)_R$ symmetry. 

In the non-Abelian case, the operator $\Ot=\d ^4\O _2$ is
non-vanishing and corresponds to infinitesimally changing $\tau$
in the Lagrangian.  To be precise, the change in the on-shell
Lagrangian upon varying $\tau \rightarrow \tau +\delta \tau$ is given
by \eqn\tauvari{\delta {\cal L}^{\rm{on-shell}}={\delta \tau \over
\ev{\tau _2}}\Ot +{\delta \overline \tau\over \ev{\tau _2}}\Otbar .}
The factor of $\ev{\tau _2 }=4\pi g_{YM}^{-2}$ in \tauvari\ is due to
the normalization of $\Ot$ given by \opnorm\ for $p=2$.  It will be
important in the next section, when we discuss modular transformation
properties.  The fact that the exactly marginal operator corresponding
to changing $\tau$ is $\sim \d ^4\Tr (\phi ^i\phi ^j)_{20'}$, where
$\Tr (\phi ^i\phi ^j)_{20'}$ is the operator $\O _2$, with $20'$ the
$SU(4)_R$ representation with Dynkin indices $(0,2,0)$, was noted in
\HST.

\newsec{Conjectures about $SL(2,Z)$ 
invariance and its bonus enhancement}

In the duality of \malda, the $SL(2,Z)$ S-duality of ${\cal N}=4$ is
tied to the $SL(2,Z)$ symmetry of $IIB$ string theory, which remains a
symmetry of the theory with $F_5$ flux and vacuum $AdS_5\times S^5$
because $F_5$ is $SL(2,Z)$ invariant.  In the supergravity limit, as
in the theory without $F_5$ flux, the $SL(2,Z)$ symmetry is enhanced
to $SL(2,R)$, with maximal compact subgroup $U(1)_Y$.  Before
discussing the bonus symmetry of the supergravity limit, we will
discuss some general ideas and speculations for how $SL(2,Z)$ acts on
correlation functions.

We expect that $SL(2,Z)$ maps any operator $\O_i$ to the {\it same}
$\O_i$ operator in the dual gauge theory, possibly up to factors to be
discussed now.  The simplest realization of the $SL(2,Z)$ invariance
of the ${\cal N}=4$ theory with $SU(N)$ gauge group (ignoring global
issues) would be that arbitrary correlation functions of operators
should be modular functions of $\tau$.  A more general possibility
would be for correlation functions to be modular forms $F^{(w,
\overline w)}(\tau, \overline \tau )$ of weights $(w,\overline w)$,
which transform as
\eqn\modform{F^{(w, \overline w)}(\tau, \overline \tau )\rightarrow 
(c\tau +d)^w(c\overline \tau +d)^{\overline w}F^{(w,\overline w)}(\tau
, \overline \tau )\qquad
\hbox{under}\qquad \tau
\rightarrow {a\tau +b\over c\tau +d}.}  One could entertain even more
general possibilities, but we will not do so here.

We expect that general correlation functions transform as \modform\
and that it is possible to assign general weights $(w_i,
\overline{w_i})$ to each operator $\O _i$.  As in \modform, $\O _i$ is
mapped under modular transformation as $\O _i\rightarrow (c\tau
+d)^{w_i}(c\overline \tau +d)^{\overline{w_i}}\O _i$ and a general
correlation function $\ev{\prod _i
\O _i (x_i)}$ will have weight $(w_T, \overline w _T)$, with $w_T=\sum
_i w_i$ and $\overline w_T=\sum _i \overline w_i$.  

Note that the factor of $g_{YM}^{-p}$ in \opnorm\ affects the weights
$(w_p, \overline w _p)$ assigned to the operator $\O _p$.  This is
because $\ev{\tau _2}\equiv 4\pi g_{YM}^{-2}$ transforms as a modular
form of weights $(-1, -1)$.  By multiplying by powers of $\ev{\tau _2}
$, it is possible to convert a modular form of weights $(w, \overline
w)$ to one of weights $(w ', \overline w'=-w')$.  We conjecture that,
with the powers of $\ev{\tau _2}$ given by the normalization condition
\opnorm, all operators $\O _i$ are modular forms of weight $(-q_i/4,
q_i/4)$, where $q_i$ is the $U(1)_Y$ charge which is assigned to the
operators.

In particular, the superconformal primary operator $\O _p$ with
normalization \opnorm\ is modular invariant.  The necessity of the
$g_{YM}^{-p}$ factor in \opnorm\ for obtaining a modular invariant
operator can be seen, for example, in the case where the gauge group
is $U(1)$ and the theory is free.  Again, this factor can be
understood as simply rescaling $\phi$ so that its kinetic term does
not have the $g_{YM}^{-2}$ factor.

To motivate the above statement about the modular weights of
descendents, consider the variation \tauvari\ of the on-shell
Lagrangian under a change $\delta \tau$ of $\tau$.  Under a modular
transformation,
\eqn\tauvarm{{\delta \tau \over \tau _2}\rightarrow \left(
{c\overline \tau +d\over c\tau +d}\right){\delta \tau \over \tau _2},}
transforms as a modular form of weight $(-1,1)$.  By assigning $\Ot$
weight $(1,-1)$, the variation \tauvari\ is modular invariant.  More
generally, operators of $U(1)_Y$ charge $q_i$ should transform with
weight $(-q_i/4, q_i/4)$.  The supercharges $Q_\alpha ^I$ and
$\overline Q_{I,\dot
\alpha }$ thus effectively transform as modular forms of weights
$(\f{1}{4}, -\f{1}{4})$ and $(-\f{1}{4}, \f{1}{4})$, respectively.  
A general correlation function thus transforms under $SL(2,Z)$ as
in \genslz.

We emphasize that the above statements apply in the ${\cal N}=4$ gauge
theory for any $N$ and $g_{YM}$ and are logically separate {}from the
AdS duality.

We turn now to the AdS duality conjecture of \malda\ and the
prescription \EW\ for computing general correlation functions: 
\eqn\EWeqn{{\cal Z}_{IIB}\left[\Phi _i|_{\partial (AdS)}=J_i(x)\right]
= \ev{e^{\sum _i \int d^4x J_i(x)\O _i(x)}}_{CFT},} for arbitrary
source functions $J_i(x)$.  In light of the above discussion, we would
like to make this prescription a bit more precise with regard to
modular transformation properties and how $\Phi _i$ is defined.
First, the $IIB$ string theory or supergravity field $\Phi _i$ must
not have any $SL(2,R)$ or $SL(2,Z)$ doublet indices $\alpha $ left
hanging loose: all should be soaked up with the $V^\alpha _\pm$
\Vmatrix.  Second, appropriate factors of $\ev{\tau _2}$ should be
introduced into the field $\Phi _i$ so that it transforms under the
modular group as a form of weights $(w_i, \overline {w_i}=-w_i)$; here
$w_i=-q_i/4$, with $q_i$ the $U(1)_Y$ charge of $\Phi _i$.  This
implies that the sources $J_i(x)$ have modular transformation
properties opposite to that of the $\O _i$ discussed above.  This
guarantees that correlation functions computed via \EWeqn\ will have
the modular transformation properties discussed above.

As a concrete example to illustrate the factors of $\ev{\tau _2}$,
consider the two point function $\ev{\O ^{(-4)}_p(x)\O ^{+4}_p(y)}$,
where $\O ^{(-4)}_p(x)\equiv \d ^4 \O _p$ and $\O ^{(+4)}_p\equiv
\overline \d ^4 \O _p$.  For $p=2$ these are 
the exactly marginal operators $\O _{p=2}^{(-4)}=\Ot$ and $\O
_{p=2}^{(+4)}=\Otbar$.  The supergravity source for $\O
^{(-4)}_{p=k+2}$ is $\delta \tau _k/\ev{\tau _2}$ and the source for
$\O ^{(+4)}_{p=k+2}$ is $\delta \overline \tau _k/\ev{\tau _2}$.  Here
$\delta \tau _{k}$ is the $k$-th $S^5$ spherical harmonic of the
variation, $\delta \tau$, of the 10d dilaton away {}from its constant
expectation value $\ev{\tau}$.  (I hope this notation for spherical
harmonics will not cause any confusion regarding $\tau _2\equiv {\rm
Im}\tau$, which is not the 2nd spherical harmonic of $\tau$.) The
reason for the factors of $1/\ev{\tau _2}$ in the source functions is,
for every spherical harmonic, it is $\delta \tau _k/\ev{\tau _2}$
which transforms with weight $\overline w=-w$: as seen by expanding
$\delta \tau$ in
\tauvarm\ in spherical harmonics, with $\tau$ set to its constant
expectation value, the $\delta \tau _k/\ev{\tau _2}$ all transform
with weight $(-1,1)$.  So $\delta \tau _{k}/\ev{\tau _2}$ and $\delta
\overline \tau _{k}/\ev{\tau _2}$ are the correct sources for the
operators $\delta ^4 \O _{p=k+2}$ and $\overline \delta ^4 \O
_{p=k+2}$, respectively, when these operators are properly normalized
as in \opnorm.

Because the total $U(1)_Y$ charge is zero, $\ev{\O
^{(-4)}_{p=k+2}(x)\O ^{+4}_{p=k+2}(y)}$ will be modular invariant.
Using \EWeqn\ with the sources as discussed above, we have
\eqn\IOWk{\ev{\O ^{(-4)}_{p=k+2}(x)\O ^{(4)}_{p=k+2}(y)}=\ev{\tau
_2}^2{\delta ^2\over \delta \tau _{k}(x)\delta \overline \tau
_{k}(y)}{\cal Z}_{IIB}[\delta \tau _k].}  The relevant
supergravity action for computing the RHS of \IOWk\ 
is simply the $k$-th $S^5$ spherical harmonic of the $S^5$
dimensional reduction of the 10d dilaton 
$\tau$ kinetic term in
\slinvact; this yields 
\eqn\FDKTk{S_{5d}={\pi ^2L^5\over 2\kappa _{10}^2}\int
d^5x\sqrt{-g}[-{1\over 2\ev{\tau _2}^2}(\partial \tau
_{k}\partial
\overline{\tau _{k}}-k(k+4)\tau _{k}
\overline{\tau _{k}})+\dots ].}
As in \FMMR, this gives  
\eqn\AIFMMR{{\delta ^2\over \delta
\tau _{k} (x)\delta \overline \tau _{k} (y)}Z_{IIB}[\delta
\tau]\sim {N^2\ev{\tau _2}^{-2}\over |x-y|^{2k+8}},} where we used
\hbaris\ but did not bother being careful with factors of $2$ and
$\pi$.  It then follows {}from \IOWk\ that 
\eqn\IOWWk{\ev{\O ^{(-4)}_p(x)\O ^{(4)}_p(y)}\sim {N^2\over
|x-y|^{2p+4}}.}  In this limit, as well as exactly, the correlation
function \IOWWk\ is independent of $\tau$, and thus modular invariant
as expected.

We now consider the enhancement of $SL(2,Z)$ to $SL(2,R)$ in the
supergravity limit of $IIB$ string theory, corresponding in the ${\cal
N}=4$ field theory to the double limit \double.  
In this limit, the supergravity source fields transform
under the full $SL(2,R)$ extension of $SL(2,Z)$, and thus the field
theory correlation functions computed via \EWeqn\ must also respect
the enlarged $SL(2,R)$ symmetry.  This means that, in this limit,
arbitrary correlation functions must transform exactly as in
\genslz, but for general $\pmatrix{a&b\cr c&d}\in SL(2,R)$, rather
than just $SL(2,Z)$. 

Because $SL(2,R)$ can be used to map any point in the upper-half plane
to any other point, its modular forms are necessarily quite trivial.
In particular, the only $SL(2,R)$ modular form which transforms as in
\modform\ with weights $\overline w=-w$ is given by
$F^{(w,-w)}=(const)\delta _{w,0}$, i.e. completely independent of
$\tau$ for $w=0$, and vanishing for $w\neq 0$.  Since correlation
functions have $\overline w=-w=q_T/4$ \genslz, we find that non-zero
correlation functions must respect the $q_T=0$, $U(1)_Y$ selection
rule \Ycons.  This is reasonable, since supergravity respects the
$U(1)_Y$ symmetry.  (As mentioned in the previous section, the
selection rule \Ycons\ is actually stronger than simple $U(1)_Y$
invariance, which would allow for non-zero net $U(1)_Y$ charge to be
soaked up by powers of $\tau$; \Ycons\ incorporates the fact that
$SL(2,R)$ invariance prevents this {}from being an option.)  Further,
the non-zero correlation functions with $q_T=0$ are independent of
$\tau$, as stated after \gencorp.

\newsec{The breaking of $SL(2,R)$ and $U(1)_Y$ in
string theory}

The tree-level worldsheet action for the $IIB$ string theory in flat
10d spacetime\foot{The worldsheet conformal field theory for the
present case of non-zero $F_5$ flux, i.e. with a Ramond-Ramond
background is not well understood. (See, however, \RMAT\ for
superstring actions argued to properly describe $AdS_5\times S^5$.)
The $F_5=0$ worldsheet theory suffices for getting insight into some
qualitative aspects, such as the breaking of $U(1)_Y$ to $Z_4$.} 
contains two terms, $S_1+S_2$ discussed in detail in sect. 5.1.2 of
\GSW.  The term $S_1$ looks well-motivated and respects the $U(1)_Y$
symmetry which rotates the two fermionic fields $\Theta$.  The term
$S_2$, looks less well-motivated but has to be added to $S_1$ to
ensure the $\kappa$ symmetry; it is independent of the worldsheet
metric and thus does not contribute to the 2d stress tensor.  The
effect of $S_2$ is also sub-leading to $S_1$ in the $\alpha '$
expansion.  The action $S_2$ violates the $U(1)_Y$ symmetry, breaking
it to $Z_4$; the $Z_4$ action involves rotating the two $\Theta$
coordinates by $\pi /2$, combined with a world-sheet parity
transformation $\sigma _1\leftrightarrow
\sigma _2$, which takes $\epsilon ^{\alpha \beta}\rightarrow -\epsilon
^{\alpha
\beta}$.  As mentioned above, in the map to ${\cal N}=4$ field theory,
this $Z_4$ corresponds to the center of the $SU(4)_R$ symmetry of the
gauge theory and thus is not an interesting new symmetry.

As discussed e.g. in \refs{\GG, \GS} and references cited
therein, the leading $\alpha '$ stringy correction to the spacetime
effective action occurs at order $(\alpha ')^3$ relative to the
supergravity effective action and has the form (in Einstein frame)
\eqn\susytens{(\alpha ')^3\int d^{10}x\sqrt{-g}(f^{(12,-12)}\lambda
^{16}+f^{(11,-11)}G\lambda ^{14}+\dots +f^{(4,-4)}G^8+\dots
f^{(0,0)}R^4 +c.c).}  The functions $f^{(w, -w)}(\tau, \overline \tau
)$ are $SL(2,Z)$ modular forms, transforming as in \modform\ with
$\overline w=-w$.  Exact expressions for the $f^{(w,-w)}$ are
conjectured e.g. in \refs{\GG, \KPoo,
\GS}, e.g. 
\eqn\ffouris{f^{(0,0)}(\tau, \overline \tau )=\sum
_{(m,n)\neq (0,0)}{\tau _2^{3/2}\over |m+\tau n|^3}.}  The expression
\ffouris\ is invariant
under $SL(2,Z)$ modular transformations, but obviously violates
$SL(2,R)$.  Although $R$ is neutral under $U(1)_Y$, the fact that
$\tau$ in \ffouris\ is charged under $U(1)_Y$ means that the $R^4$
terms in \susytens\ also violates $U(1)_Y$ (though clearly preserves
the $Z_4$ since $\tau$ has charge 4), as do the other terms in
\susytens\ more explicitly.  

As in \BG, assuming that the duality of \adsr\ applies away {}from the
supergravity limit, with the sub-leading
stringy terms in
\ffouris, leads to predictions for the sub-leading corrections to 
the ${\cal N}=4$ field theory correlation functions away {}from the
double limit.  Using \susytens, we find
\eqn\corsti{\ev{\prod _{i=1}^n \O ^{(q_i)}_i (x_i )}=
N^2f^{(0)}_{i_1\dots
i_n}(x_i)\delta _{q_T,0}+N^{1/2}f^{(-q_T/4,q_T/4)}(\tau, \overline
\tau )f^{(3)}_{i_1\dots i_n}(x_i)+\dots,} where $q_T=\sum _i q_i$ is
the total $U(1)_Y$ charge of the operators (which is opposite to that
of the supergravity source fields).  Here 
$f^{(0)}_{i_1\dots i_n}(x_i)$ is
the leading supergravity contribution and $f^{(3)} _{i_1\dots
i_n}(x_i)$ are the leading corrections to supergravity amplitudes
coming {}from the additional interactions in \susytens.  The relative
factor of $N^{-3/2}$ in \corsti\ comes {}from the $(\alpha ')^3$ in
\susytens, along with \strexp.  The modular forms in \corsti\ are the
same ones appearing in \susytens, e.g. $f^{(12,-12)}(\tau, \overline
\tau)$ for the 16-point function of the operator $\delta ^3 \O _p$,
of $U(1)_Y$ charge $Y=-3$, which is conjugate to the
supergravity source $\lambda$.  

The fact that the modular forms in \corsti\ have weights $(-q_T/4,
q_T/4)$ is seen in \susytens: the weights of the modular forms are
correlated in this way with the $U(1)_Y$ charge of the interaction
terms in \susytens.  This means that the corrections in
\corsti\ respect the $SL(2,Z)$ symmetry with our conjectured 
general modular transformation
property \genslz.

The stringy correction term in \corsti\ gives the leading correction,
away {}from the double limit, which violates the approximate bonus
$SL(2,R)$ and $U(1)_Y$ symmetries of correlation functions.  It is
subleading by $N^{-3/2}$ for any fixed $g_{YM}$.  In the small
$g_{YM}$ limit, the leading contributions to the modular forms in
\susytens\ occur at string tree-level and are
$f^{(-q_T/4,q_T/4)}=(const)g_{YM}^{-3/2} +\dots$.  In this limit, we
see {}from
\corsti\ that the violations of the bonus symmetries are subleading by
order $(g_{YM}^2N)^{-3/2}$, as expected {}from
\strexp\ (in the limit of small $g_{YM}$, $D$-string effects of size
\strexpp\ can be ignored).   There are also terms in the small
$g_{YM}$  expansion of the $f^{(-q_T/4,q_T/4)}$ which correspond to 
Yang-Mills instanton contributions to the correlation functions
\corsti.  It was argued in \BGKR\ and, more recently extensively
analyzed and verified in \DHKMV, that $SU(N)$ Yang-Mills instantons do
lead to contributions to correlation functions precisely as expected
{}from \corsti, with precisely the same instanton coefficients as
obtained by expanding the $f^{(-q_T/4, q_T/4)}$.  Violations of the
bonus symmetries which do not get a contribution {}from the $(\alpha
')^3$ terms in \susytens\ are even more sub-leading in the
$(g_{YM}^2N)^{-1}$ expansion.

In \BG\ it was pointed out that the $R^4$ term does not contribute to
$n<4$ point functions of the stress tensor because
\eqn\drdg{{\delta ^n\over \delta g^n}R^4|_{AdS_5\times S^5}=0\quad{\rm
for}\quad n=0,1,2,3.} Similarly, the other terms in \susytens\ and low
numbers of variations with respect to the fields also vanish when
evaluated for the $AdS_5\times S^5$ vacuum.  The first non-zero
contribution {}from \susytens\ is that of \BG, where the $R^4$ term
contributes to the four-point function $\ev{\prod _{i=1}^4T_{\mu _i\nu
_i}(x_i)}$.  This leads to violation of the $SL(2,R)$ symmetry
starting at four-point functions.  Using \corrd, the $\tau$ dependence
of this term also leads to violation of the $U(1)_Y$ selection rule
starting at the 5-point function $\ev{\Ot (z)\prod _{i=1}^4T_{\mu
_i\nu _i}(x_i)}$.  The other $U(1)_Y$ violating terms in \susytens\
are only non-vanishing for higher $n$-point functions, e.g. the $G^8$
term for $n=8$ point functions.

While \susytens\ is just the leading string correction in the $\alpha
'$ expansion, we expect that, via the arguments of \KR, all higher
order $\alpha '$ corrections to the effective supergravity action will
also have the property, as in \drdg, that they vanish when evaluated
for low numbers of variations around the $AdS_5\times S^5$ vacuum.  We
thus expect that $SL(2,R)$ and $U(1)_Y$ are actually exact symmetries
of $n\leq 3$ point functions for all $g_{YM}$ and $N$.  The $SL(2,R)$
symmetry of $n\leq 3$ point functions is the conjecture of
\LMRS\ that these correlation functions are independent of $g_{YM}$
(along with $\theta _{YM}$) for finite $N$.  Descendant $n\leq
3$-point correlation functions will also be independent of $g_{YM}$
and $\theta$ for finite $N$, and respect the $U(1)_Y$ selection rule
\Ycons\ exactly.  The cancellations of radiative corrections exhibited
in \DFS\ support these conjectures. 

We make a slightly stronger conjecture, which is suggested by
 \susytens: that the $U(1)_Y$ selection rule is an exact selection
rule for all $n\leq 4$-point functions.  Using \corrd, this implies
that all $n\leq 3$-point correlation functions are independent of
$\tau$.  

The non-trivial $N$ dependence of the $n\leq 3$-point functions
discussed in \DFS\ must correspond, via \hbaris, to non-trivial string
loop corrections to these amplitudes.  As mentioned in \LMRS, one
might expect that the scattering of three gravitons is not affected by
quantum corrections.  We note that this is actually completely
consistent with the normalization of the 3-point function of the
massless $\O _2$ multiplet, which includes the conserved currents, if
\hbaris\ is simply modified to $\hbar \sim (N^2-1)^{-1}$ for gauge
group $SU(N)$ rather than $U(N)$.  This can be understood simply as a
one-loop string correction to the relation between $\kappa _{10}$ and
$\kappa _5$ by $S^5$ dimensional reduction.

\newsec{Proving exact $U(1)_Y$ 
invariance of $n$-point functions for low
$n$.}

We will now prove that all two-point functions of operators in short
representations respect the $U(1)_Y$ selection rule \Ycons\ for all
$g_{YM}$, $\theta _{YM}$, and $N$.  Note that this selection rule is
not a trivial consequence of the $SU(2,2)\times SU(4)_R$ symmetry, as
there are two-point functions which would respect these symmetries but
violate $U(1)_Y$ if they were non-zero.  For example, a non-zero
two-point function of the operator of the form $\d ^4\O _p$, which is
a Lorentz scalar and in the $(0,p-2,0)$ representation of $SU(4)_R$,
with $U(1)_Y$ charge $-4$, with itself would respect $SU(2,2)\times
SU(4)_R$ but violate $U(1)_Y$; our argument shows that this and all
other $U(1)_Y$ violating two-point functions vanish.

Consider the correlation functions in Euclidean space, with radial
ordering {}from the origin (an arbitrary point).  We then have vacuum
states $|0\rangle$ and $\langle 0|$, which are annihilated by all
supercharges, and correlation functions are to be understood as: 
$\langle 0|\prod _i \O _i
(x_i) |0\rangle$.  For arbitrary operators $A$ and $B$,
\eqn\ABQ{\langle 0|[Q,A(x)]_{\pm}B(y)|0\rangle =\pm \langle
0|A(x)QB(y)|0\rangle =\pm \langle 0|A(x)[Q,B(y)]_{\pm}|0\rangle,}
since $Q$ annihilates $\langle 0|$ and $|0\rangle$; the same identity
holds with $Q$ replaced by $\overline Q$.  By repeating the operation
\ABQ, an arbitrary two-point function, of any operators in the table
in appendix A, is equal to a two-point function of the form
\eqn\arbtpf{\ev{\O_{p}(x)[D^{(n,\overline n)}\O_{q}](y)},}
where $\O_p(x)$ is the superconformal primary scalar operator with
dimension $\Delta =p$ and $SU(4)_R$ representation $(0,p,0)$ and
$[D^{(n, \overline n)}\O_q](y)$ is a superconformal descendent
obtained by acting with $n$ $Q$ and $\overline n$ $\overline Q$
operators on $\O_q(y)$.  The two-point function violates $U(1)_Y$ if
it is non-zero for $n\neq \overline n$.

For two-point functions, there is an essential difference between
whether the superconformal descendent $[D^{(n, \overline n)}\O _q]$ is
a primary field or a descendent under the conformal group $SU(2,2)$.
The superconformal descendents in the table in appendix A are all
primary under the conformal group $SU(2,2)$.  Each has an infinite
tower of conformal descendents obtained by acting with $P_\mu$,
corresponding to taking $x_\mu$ derivatives of the operator.  As is
well known, using the Ward identities of the $SU(2,2)$ conformal
group, it can be shown that the two-point function of two conformal
primary operators can be non-zero only if their conformal dimensions
are equal.  Thus, if $[D^{(n, \overline n)}\O _q]$ is a $SU(2,2)$
conformal primary operator, the two-point function \arbtpf\ can be
non-zero only if it has dimension $\Delta =p$ and in an $SU(4)_R$
representation which includes a singlet in its product with the
$(0,p,0)$ representation of $\O _p$.  The only operator in the table
in appendix A which has these properties is the superconformal primary
operator $\O _p$ itself, i.e. $n=\overline n=0$.  The two-point
function of $\O _p$ with itself of course respects $U(1)_Y$, as $\O
_p$ is neutral.

Thus any two-point function involving a superconformal descendant
which could potentially violate the $U(1)_Y$ selection rule will
vanish unless, upon using \ABQ\ to write it in the form \arbtpf, the
operator $[D^{(n, \overline n)}\O _q]$ is {\it not} a $SU(2,2)$
conformal primary operator.  This can happen because, using the
supersymmetry algebra, a $Q$ and $\overline Q$ anti-commutator is
replaced with $P_\mu$.  The $P_\mu$ can be replaced with $\partial
_{y_\mu}$ acting on the two-point function for the remaining
operators, which is again of the form \arbtpf\ but with an operator of
the form $[D^{(n-1, \overline n-1)}\O _q](y)$, since a $Q$ and
$\overline Q$ were traded for the $\partial _{y_\mu}$.  Repeating the
above argument, the two-point function on which $\partial _{y_\mu}$
acts can also only be non-zero if $(n-1, \overline n-1)=0$ and $q=p$
or if $[D^{(n-1,n-1)}\O _q]$ is a $SU(2,2)$ descendent,
$[D^{(n-1,n-1)}\O _q]=[P_{\mu '}, [D^{(n-2,n-2)}\O _q]]$.  Continuing
this argument, the only non-zero two-point functions have in \arbtpf\
$n=\overline n$ and $q=p$, with $[D^{(n,n)}\O _q]$ a $SU(2,2)$
descendent of $\O _p$.

Thus all non-zero two-point functions respect the $U(1)_Y$ selection
rule and can be written as $y_\mu$ derivatives of the two-point
function of superconformal primary operators $\ev{\O _p(x)\O _p(y)}$.
For example, $\ev{T_{\mu \nu}(x)T_{\rho \sigma}(y)}$ and $\ev{\Ot
(x)\Otbar (y)}$ can each be written as particular combinations of four
$\partial _y$ derivatives acting on $\ev{\O _2(x)\O _2(y)}$, while
$\ev{\Ot \Ot}=0$.  All two-point functions of superconformal
descendents and, in particular, their normalization, are fixed by the
primary $\ev{\O _p(x)\O _p (y)}$ correlation functions.  This
analysis, again, is valid for all $g_{YM}$, $\theta _{YM}$, and $N$.

We note that, because all two-point functions exactly respect the
$U(1)_Y$ selection rule, the Born-approximation calculation of an
arbitrary $n$-point function, where the $n$-point function is broken
up into products of two-point functions, will also respect $U(1)_Y$.
This approximation gives the leading contribution to the correlation
function in the small $g_{YM}^2N$ limit.  Thus arbitrary
$n$-point correlation functions will also respect the $U(1)_Y$
selection rule in the small $g_{YM}^2N$ limit:
\eqn\arbborn{\ev{\prod _{i=1}^n\O ^{(q_i)}_i(x_i)}=F_{i_1\dots
i_n}(x_i;N)\delta _{q_T,0}+\rm{order}\ (g_{YM}^2N),} 
with $q_T=\sum _{i=1}^nq_i$ and $F_{i_1\dots i_n}(x_i;N)$ independent
of $g_{YM}$ and $\theta_{YM}$.  This is valid for arbitrary $N$ and,
in the limit of large $N$, 
\eqn\Fapprox{F_{i_1\dots i_n}(x_i;N)\approx
N^2H_{i_1\dots i_n}(x_i),} where the functions $H_{i_1\dots i_n}$
could generally differ {}from those of \gencorp, which described the
large $g_{YM}^2N$ limit, as arbitrary correlation functions generally
depend on $g_{YM}^2N$.  It would be interesting to check the field
theory prediction \arbborn\ against $IIB$ string theory and the
duality of \adsr: stringy violations of $U(1)_Y$ must also vanish in
the {\it small} $\lambda =g_{YM}^2N$ limit.  Checking this in string
theory would require better understanding of the worldsheet CFT with
non-zero $F_5$ flux.

Manipulations of the type used above do not seem as useful for higher
$n$-point functions.  Although we expect that the $U(1)_Y$ symmetry is
an exact symmetry for three-point functions and possibly also
four-point functions, we have here succeeded only in proving it for
two-point functions.  In the next section, we discuss a formalism
which {\it should} just be a convenient way to re-package the
superconformal Ward identities.  As we will discuss, however, this
formalism is extremely powerful -- perhaps too powerful!

\newsec{Harmonic superspace and the $U(1)_Y$ symmetry}

The ${\cal N}=4$ gauge superfield $W$, as well as the operators in
small representations, obey constraints which imply that they only
depend on half of the coordinates of a would-be superspace.  This is
seen in the table, in that the small representations truncate at
$\d ^4\overline \d ^4\O _p$ rather than $\d ^8\overline \d ^8\O _p$.  
It is impossible to implement this constraint in superspace in which
$SU(4)_R$ is manifest.  Introducing Grassmann coordinates $\Theta
^I_\alpha$ and $\overline \Theta _{I, \dot \alpha}$ conjugate to
$Q^I_\alpha$ and $\overline Q_{I,\dot \alpha}$, the gauge superfield
should depend on two of the four possible $\Theta _\alpha$ coordinates
and two of the four possible $\overline \Theta _{\dot \alpha}$
coordinates.  Thus at most a $SU(2)\times SU(2)$ subgroup of $SU(4)_R$
can be made manifest.  Basically, we decompose the supersymmetries
under $SU(4)_R\rightarrow SU(2)\times SU(2)\times U(1)$ as ${\bf
4}\rightarrow {\bf (2,1)_1\oplus (1,2)_{-1}}$ and ${\bf \overline
4}\rightarrow {\bf (\overline 2,1)_{-1}\oplus (1,\overline 2) _1}$ and
only keep fermionic coordinates for ${\bf (2,1)_1}$ and ${\bf (1,
\overline 2)_1}$.

The result, then, is $N=4$ harmonic (or
``analytic'') superspace, involving coordinates
\eqn\analytic{X=\pmatrix{x_{\alpha \dot \alpha} &\lambda _{\alpha
a'}\cr \pi _{a\dot \alpha} &y_{aa'}},} where $\lambda _{\alpha a'}$
and $\pi _{a\dot \alpha}$ are the fermionic coordinates, with $a=1,2$
and $a'=1,2$ labels for the $SU(2)\times SU(2)'\subset SU(4)_R$ and
$y_{aa'}$ a bosonic coordinate living on the Grassmanian coset space
$SU(4)/S(U(2)\times U(2)')$.  Because this coset space is compact, the
power of $y_{aa'}$ is constrained. The $y_{aa'}$ coordinates allow the
$SU(2)\times SU(2)'$ indices to be converted back to $SU(4)_R$ indices
at the end of the day.  This formalism has been discussed in detail in
the series of papers \refs{\harmrefs, \HWinvts}.
 
This superspace formalism is a remarkably powerful technology: it
allows the ${\cal N}=4$ gauge multiplet to be packaged into a single
superfield $W(X)$, and the entire collection of small ${\cal N}=4$
representation operators appearing in the table of appendix $A$ to be
neatly packaged into the single super-space operator $A_p(X)=\Tr
_{SU(N)} W(X)^p$.  General correlation functions of small
representation operators then take the form
\eqn\corrform{\ev{\prod _{i=1}^nA_{p_i}(X_i)}=f_{p_1,\dots p_n}(X_1,
\dots X_n;N; \tau).}
A key dynamical assumption \harmrefs\ is that the function
$f_{p_i}(X_i)$ remains analytic, with only positive powers of
$y_{aa'}$, in the quantum theory.  In the last reference of \harmrefs,
this principle was explicitly checked for a particular correlation
function, at the two loop level, in the analogous ${\cal N}=2$
harmonic superspace formalism; all intermediate non-harmonic analytic
terms were found to cancel upon adding contributions {}from all
diagrams.  

The function in \corrform\ is then constrained by superconformal
invariance; we now summarize the results of \refs{\harmrefs,
\HWinvts}. Two-point functions and three-point functions are argued to
be completely fixed to be
\eqn\twop{\ev{A_{p}(X_1)A_{q}(X_2)}=c_{p}\delta _{p,q}g_{12}^p,}
\eqn\threep{\ev{A_{p_1}A_{p_2}A_{p_3}}=c_{p_1p_2p_3}(g_{12})^{\half 
(p_1+p_2-p_3)}(g_{23})^{\half (p_2+p_3-p_1)}(g_{13})^{\half
(p_1+p_3-p_2)},} 
where $c_p$ and $c_{p_1p_2p_3}$ are (a priori, 
possibly $\tau$ dependent)
constants and
\eqn\gijis{g_{ij}\equiv 
(\sdet X_{ij})^{-1}={\widehat y_{ij}^2\over
x_{ij}^2},}
\eqn\yis{\widehat{(y_{ij})}_{aa'}=(y_{ij})_{aa'}-{(\pi _{ij})_{a\dot
\alpha} (x_{ij})^{\dot \alpha \alpha}(\lambda _{ij})_{\alpha a'}\over 
(x_{ij})^2},} and $X_{ij}\equiv X_i-X_j$ (i.e. $y_{ij}=y_i-y_j$ etc.).
$n$-point functions with $n\geq 4$ again involve the $g_{ij}$ \gijis,
though now there can also be undetermined functions of superconformal
invariants:
\eqn\Fgen{\ev{\prod _{i=1}^nA _{p_i}(X_i)}=\prod
_{i<j}g_{ij}^{(p_i+p_j-{p_T\over n-1})/(n-2)}F_{p_i}(I),}
where $p_T=\sum _{i=1}^np_i$ and $I$ are all possible superconformal
invariants. The possible superconformal invariants were classified in
\HWinvts\ and found to be of two types.  The first are
super-cross-ratios of the $g_{ij}$:
\eqn\scrossr{{g_{ij}g_{kl}\over g_{ik}g_{jl}}.}
The second type of superconformal invariants involve super-traces $str
N_i^p$, with $p=1,\dots 4$, of quantities $N_i$ defined in \HWinvts,
the simplest example, for four points, being
\eqn\streg{str N=str(X_{12}^{-1}X_{23}X_{34}^{-1}X_{41}).}
As remarked in \refs{\harmrefs, \HWinvts}, the condition that there be
no $y_{ij}$ singularities puts constraints on the dependence of this
second class of invariants; these aspects will not be relevant for the
point we are making here.

Having described this powerful formalism, it must be mentioned that
its applicability is considered suspicious by some physicists.  (See,
for example, in the discussion of descendent correlation functions in
\DFS.)  A reason for concern is that there is no known off-shell
superspace for ${\cal N}=4$ supersymmetry \NOS; the present formalism
is purely on-shell.  The danger, then, is that it is incapable of
reproducing the off-shell contributions to correlation functions in
intermediate channels.

We will argue that assuming applicability of this formalism leads to
an incorrect result: all correlation functions of operators in short
representations of the superconformal group would {\it exactly}
respect the $U(1)_Y$ selection rule!  If correct, this would imply, as
a consequence of \corrd, that {\it all} correlation functions of
operators in short multiplets are completely independent of $g_{YM}$.
However, as discussed in footnote 2, this latter result has been shown
to be {\it incorrect}, as $n\geq 4$-point functions are definitely 
renormalized.

To see the above result about $U(1)_Y$, note that $U(1)_Y$ charge in
this formalism is carried by $\lambda _{\alpha a'}$, which has charge
$+1$, and $\pi _{a\dot \alpha}$, which has charge $-1$.  The bosonic
coordinates $x_{\alpha \dot \alpha}$ and $y_{aa'}$ are, of course,
neutral under $U(1)_Y$.  In order to have a correlation function which
does not respect the $U(1)_Y$ symmetry, the RHS of \corrform\ would
have to contain a function $f_{p_i}(X_i; N; \tau)$ which is not
invariant under the $U(1)_Y$ transformation
\eqn\pilt{(\lambda _i)_{\alpha a'}\rightarrow 
C (\lambda _i)_{\alpha a'} \qquad\rm{and}\qquad (\pi _i)_{a\dot \alpha}
\rightarrow C^{-1}(\pi _i)_{a\dot \alpha},} 
for an arbitrary phase $C$ (which could just as well be an arbitrary
complex number, corresponding to $U(1)_Y$ complexified).  This
transformation can be represented on the $X_i$ coordinates \analytic\
as 
\eqn\Tis{X_i\rightarrow T^{-1}X_iT, \qquad \rm{with}\quad 
T=\pmatrix{\sqrt{C}1_2&0\cr 0&1_2}\in GL(2|2).}  Since $\sdet T=C$,
this $T$ is not in $SL(2|2)$ for a non-trivial $U(1)_Y$
transformation.

It is easily seen {}from \gijis\ and \yis\ that
the $g_{ij}$ are invariant under the $U(1)_Y$ transformation \pilt\ or
\Tis.  Upon expanding out both sides of
\twop\ and \threep\ in components, it then follows that all two point
and three point functions of operators with non-zero total $U(1)_Y$
charge necessarily vanish.  These results are plausible and in line
with our conjecture, and with the descendent 3-point function
calculation in \DFS, which had non-zero net $U(1)_Y$ charge and was
found to vanish to leading and next-to-leading order in a small
coupling expansion.

Moving on to four and higher point functions, the $g_{ij}$ terms in
\Fgen, again, respect the $U(1)_Y$ selection rule.  Thus the only way
there could be terms on the right side of \Fgen\ with non-zero
$U(1)_Y$ charge is if some of the superconformal invariants $I$ carry
non-zero $U(1)_Y$ charge.  It is clear that all invariants of the
first type \scrossr\ respect $U(1)_Y$, since the $g_{ij}$ all respect
$U(1)_Y$.  Further, the invariants of the second type also respect
$U(1)_Y$.  Clearly \streg\ is invariant under \Tis.  Indeed, the
transformation \pilt\ is achieved in terms of the $u_i=(1,\ X_i)$
coordinates of \HWinvts\ by $u_i\rightarrow T^{-1}u_i g_T$, with $g_T
=diag (T,T)$, with $T$ given by \Tis.  $g_T$ is in $GL(4|4)$ rather
than $SL(4|4)$, but the basic superconformal ingredients $K_i$ and
$L_i$ defined in eqns. (27) and (28) of \HWinvts\ are clearly
invariant under $u_i\rightarrow u_ig_T$ anyway.  The final invariants,
by construction, must also be invariant under the $u_i\rightarrow
T^{-1}u_i$ transformation needed to take $u_i$ back to the form
$(1,X')$.  Thus all invariants constructed in
\HWinvts\ respect the $U(1)_Y$ symmetry.

We thus obtain a result which is incorrect: that, for all $g_{YM}$,
and $N$, all $n$-point correlation functions of short representation
operators exactly obey the exact $U(1)_Y$ selection rule \Ycons, which
would imply their non-renormalization.  This is contrary to the
results of \refs{\BGKR, \DZF, \GRPS, \EHSSW, \DHKMV}, where it was
explicitly shown that various $n\geq 4$ point functions do, in fact,
get renormalized.  Again, we have conjectured that the $U(1)_Y$
selection rule actually is exact for $n\leq 4$ point functions, which
would imply non-renormalization only for $n\leq 3$ point functions.

There are two options at this juncture:

\lfm{(1)} The ${\cal N}=4$ harmonic superspace formalism is 
inherently problematic.  Again, this might have been expected as it is
a purely on-shell formalism.  

\lfm{(2)} The ${\cal N}=4$ harmonic superspace formalism can be
salvaged by finding some new superconformal invariants, which violate
$U(1)_Y$, which have been overlooked in the classification of
\HWinvts.  This would allow the above incorrect 
conclusions about the exact $U(1)_Y$ selection rule to be evaded.

\noindent

Option (1) would be unfortunate.  

It would be nicest if option (2) is correct and that, in line with our
conjecture, there is (at least one) as-yet missing superconformal
invariant, which violates $U(1)_Y$, and which can only be written down
for $n>4$ point functions.  However, I have not yet succeeded in
constructing such an invariant.  Again, this issue in no way affects
the results and conjectures of the previous sections.

\bigskip
\centerline{{\bf Acknowledgments}}

I would like to thank N. Seiberg, W. Skiba, and E. Witten for useful
discussions.  I would also like to thank the anonymous referee for
his or her careful reading of the manuscript, and the many helpful 
suggestions for improving the presentation and references.  
This work was supported by UCSD grant DOE-FG03-97ER40546
and an Alfred Sloan Foundation Fellowship.  The final stages of this
work, at the IAS, was also supported by the W.M. Keck Foundation.

\bigskip

\vfill
\eject
\appendix{A}{Table of the spectrum of short multiplets}
$$\matrix{{\rm form} &&SO(4)&&\Delta && SU(4)_R&&
U(1)_Y\cr 
\O _p&&(0,0)&&p&&(0,p,0)&&0\cr
\delta \O _p&&(\half,0)&&p+\half &&(0,p-1,1)&&-1\cr
\overline \delta\O _p&&(0,\half)&&p+\half &&(1,p-1,0)&&1\cr
\delta^2\O _p&&(1,0)&&p+1&&(0,p-1,0)&&-2\cr
\overline \d ^2\O _p&&(0,1)&&p+1&&(0,p-1,0)&&2\cr
\d ^2\O _p&&(0,0)&&p+1&&(0,p-2,2)&&-2\cr
\overline \d ^2\O _p&&(0,0)&&p+1&&(2,p-2,0)&&2\cr
\d\overline \d\O _p&&(\half, \half )&&p+1&&(1,p-2,1)&&0\cr
\d ^3\O _p&&(\half, 0)&&p+\f{3}{2}&&(0,p-2,1)&&-3\cr
\overline \d ^3\O _p&&(0,\half )&&p+\f{3}{2}&&(1,p-2,0)&&3\cr
\d ^2\overline \d\O _p&&(1,\half)&&p+\f{3}{2}&&(1,p-2,0)&&-1\cr
\overline \d ^2\d \O _p&&(\half,1)&&p+\f{3}{2}&&(0,p-2,1)&&1\cr
\d ^4\O _p&&(0,0)&&p+2&&(0,p-2,0)&&-4\cr
\overline \d ^4\O _p&&(0,0)&&p+2&&(0,p-2,0)&&4\cr
\d ^2\overline \d ^2\O _p&&(1,1)&&p+2&&(0,p-2,0)&&0\cr
\d \overline \d ^2\O _p&&(\half, 0)&&p+\f{3}{2}&&(2,p-3,1)&&1\cr
\overline \d \d ^2\O _p&&(0,\half )&&p+\f{3}{2}&&(1,p-3,2)&&-1\cr
\d ^3\overline \d \O _p&&(\half, \half )&&p+2&&(1,p-3,1)&&-1\cr
\overline \d ^3\d \O _p&&(\half, \half )&&p+2&&(1,p-3,1)&&1\cr
\d ^2\overline \d ^2\O _p&&(1,0)&&p+2&&(2,p-3,0)&&0\cr
\d ^2\overline \d ^2\O _p&&(0,1)&&p+2&&(0,p-3,2)&&0\cr
\d \overline \d ^4\O _p&&(\half, 0)&&p+\f{5}{2}&&(0,p-3,1)&&3\cr
\overline \d  \d ^4\O _p&&(0,\half )&&p+\f{5}{2}&&(1,p-3,0)&&-3\cr
\d ^2\overline \d ^3\O _p&&(1,\half )&&p+\f{5}{2}&&(1,p-3,0)&&1\cr
\overline \d ^2\d ^3\O _p&&(\half , 1)&&p+\f{5}{2}&&(0,p-3,1)&&-1\cr
\d ^2\overline \d ^4\O _p&&(1,0)&&p+3&&(0,p-3,0)&&2\cr
\overline \d ^2\d ^4\O _p&&(0,1)&&p+3&&(0,p-3,0)&&-2\cr
\d ^2\overline \d ^2\O _p&&(0,0)&&p+2&&(2,p-4,2)&&0\cr
\d ^3\overline \d ^2\O _p&&(\half, 0)&&p+\f{5}{2} &&(2,p-4,1)&&-1\cr
\overline \d ^3\d ^2\O _p&&(0,\half )&&p+\f{5}{2}&& (1,p-4,2)&&1\cr
\d ^2\overline \d ^4\O _p&&(0,0)&&p+3&&(0,p-4,2)&&2\cr
\overline \d ^2 \d ^4\O _p&&(0,0)&&p+3&&(2,p-4,0)&&-2\cr
\d ^3\overline \d ^3\O _p&&(\half, \half) &&p+3&&(1,p-4,1)&&0\cr
\d ^3\overline \d ^4\O _p&&(\half, 0)&&p+\f{7}{2} && (0,p-4,1)&&1\cr
\overline \d ^3\d ^4\O _p&&(0,\half )&&p+\f{7}{2}&&(1,p-4,0)&&-1\cr
\d ^4\overline \d^4\O _p&&(0,0)&&p+4&&(0,p-4,0)&&0\cr}$$

\listrefs \end